          [2001/04/25 1.1 (PWD)]
\documentclass[a4paper,twocolumn]{esapub2005}
\usepackage{times}
\usepackage{natbib}
\usepackage{epsfig}

\title{Search for X-ray bursts in the INTEGRAL/IBIS data of
2003-2005 and\protect\\ discovery of the new X-ray burster 
IGR J17364-2711/17380-3749}

\author[1]{I.V. Chelovekov}
\author[1]{S.A. Grebenev}
\author[1,2]{R.A. Sunyaev}
\affil[1]{Space Research Institute, Russian Academy of Sciences,
Profsoyuznaya 84/32, 117997 Moscow, Russia} 
\affil[2]{Max-Planck-Institut f\"ur Astrophysik, 
Karl-Schwarzschild-Str. 1, 85740 Garching bei M\"unchen, Germany}

\def\deg{\hbox{$^\circ$}}
\def\la{\mathrel{\hbox{\rlap{\hbox{\lower4pt\hbox{$\sim$}}}\hbox{$<$}}}}
\def\ga{\mathrel{\hbox{\rlap{\hbox{\lower4pt\hbox{$\sim$}}}\hbox{$>$}}}}
\def\uh{\hbox{$^{\rm h}$}}
\def\um{\hbox{$^{\rm m}$}}
\def\us{\hbox{$^{\rm s}$}}
\def\arcmin{\hbox{$^\prime$}}
\def\arcsec{\hbox{$^{\prime\prime}$}}
\def\ergs{erg~s$^{-1}$}

\begin{document}


\maketitle

\begin{abstract}
All the observations performed with the IBIS telescope aboard
the INTEGRAL observatory during the first 2.5 years of its
in-orbit operation have been analyzed to find X-ray
bursts. There were 1788 statistically confident 
events
with a duration from 5 to 500 s
revealed in time records of the 15--25 keV count rate of the
IBIS/ISGRI detector, 319 of them were localized and, with one
exception, identified with persistent X-ray sources. The known
bursters were responsible for 215 of the localized events. One
burst$\,$was$\,$detected$\,$from$\,$AXJ1754.2-2754, the source previously
unknown as a burster, and another burst --- from a new source.
There was duality in determining its
position --- its name could be either IGR~J17364-2711 or
IGR~J17380-3749. Curiously enough, the 138 bursts were detected
from one X-ray burster ---
\mbox{GX 354-0}.
\end{abstract}

\section{Introduction}

Many X-ray bursts detected by orbital observatories are 
associated with thermonuclear explosions on weakly magnetized
accreting neutron stars (type-I bursts). They provide us with direct
information on processes near the surface of neutron stars
under conditions of strong gravity, extreme pressure and high
temperatures. The detection of type-I bursts, along with the
detection of coherent pulsations, serves the most
important criterion for identifying the nature of
the compact object in X-ray binaries.

Type-I bursts are generally observed from weak X-ray sources (or
transients in their quiescent state). During the burst the
luminosity of such a source (a burster) can increase by two or
three orders of magnitude relative to its persistent level,
reaching a critical Eddington value. The typical burst at the
Galactic center distance ($\sim8.5$ kpc) provides a peak flux of
about 2--3 Crab and can be easily detected. This opens up a
possibility for using the bursts in searching for previously
unknown bursters with persistent X-ray fluxes below the level of
reliable detection by currently available wide-field 
instruments.

The INTEGRAL observatory \citep{cw03} is best suited for
performing such a search. It is equipped with unique wide-field
telescopes that allow sky fields with an area of $\sim$ 1000
square degrees to be simultaneously studied with a flux
sensitivity up to 1 mCrab (over several hours of
observations) and an angular resolution reaching several
arcminutes. In addition, INTEGRAL devotes up to 85\% of the
whole time to observations of the Galactic center region and the
Galactic plane, where the bulk of the Galactic stellar mass is
concentrated.

In this paper, to find X-ray bursts, we analyze time records of
the total 15-25 keV count rate of the IBIS/ISGRI detector
\citep{fl03,pu03} aboard INTEGRAL obtained 
from February 10, 2003, through August 31, 2005.
The main objective of this study was an attempt to discover
new bursters or fast X-ray transients. Some of the presented
results (corresponding to the observations before July 2, 2004)
have already been published in
\citep{chelovekov06}.

\section{\bf OBSERVATIONS AND DATA ANALYSIS}

\begin{figure}[hb]
\centering
\epsfig{file=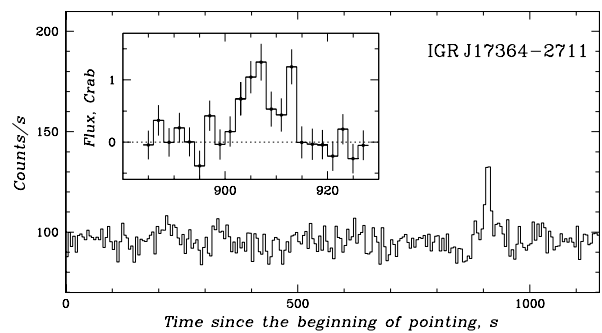,width=0.98\linewidth}
\caption{IBIS/ISGRI count rate in the 15-25 
keV band recorded on February 17, 2004, during the pointing when an
X-ray burst from a previously unknown burster was discovered
(the insertion shows its reconstructed profile).\label{fig1}}
\end{figure}

Our analysis of the IBIS data was based on
the OSA 4.2 software package. 
Using the list of events from its GTI task, we reproduced time
records of the count rate (see Fig.\,\ref{fig1}) for each individual INTEGRAL
observation (cor-\\ [-2mm]
\begin{figure*}[t]
\centering
\epsfig{file=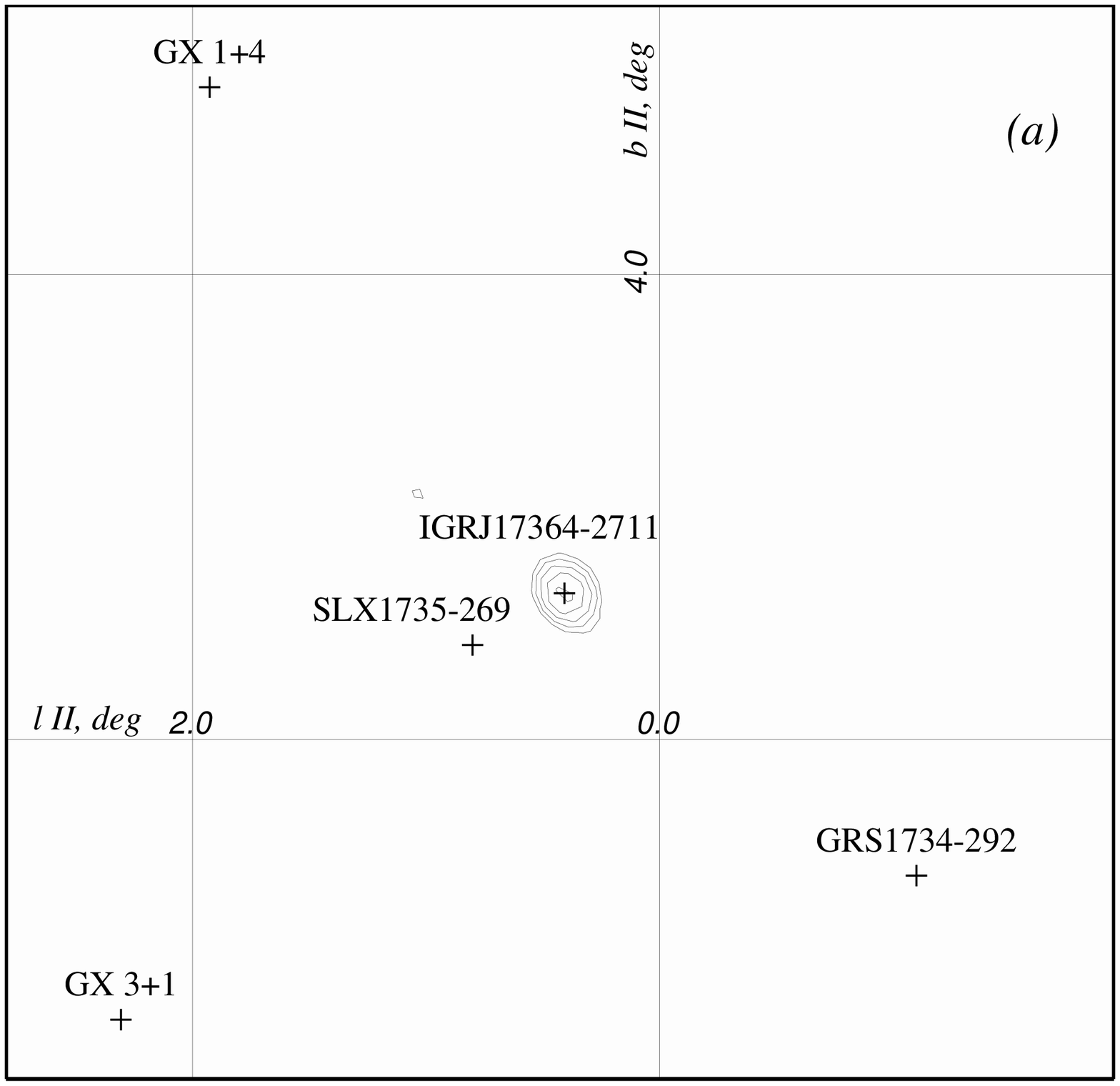,width=0.33\linewidth}
\epsfig{file=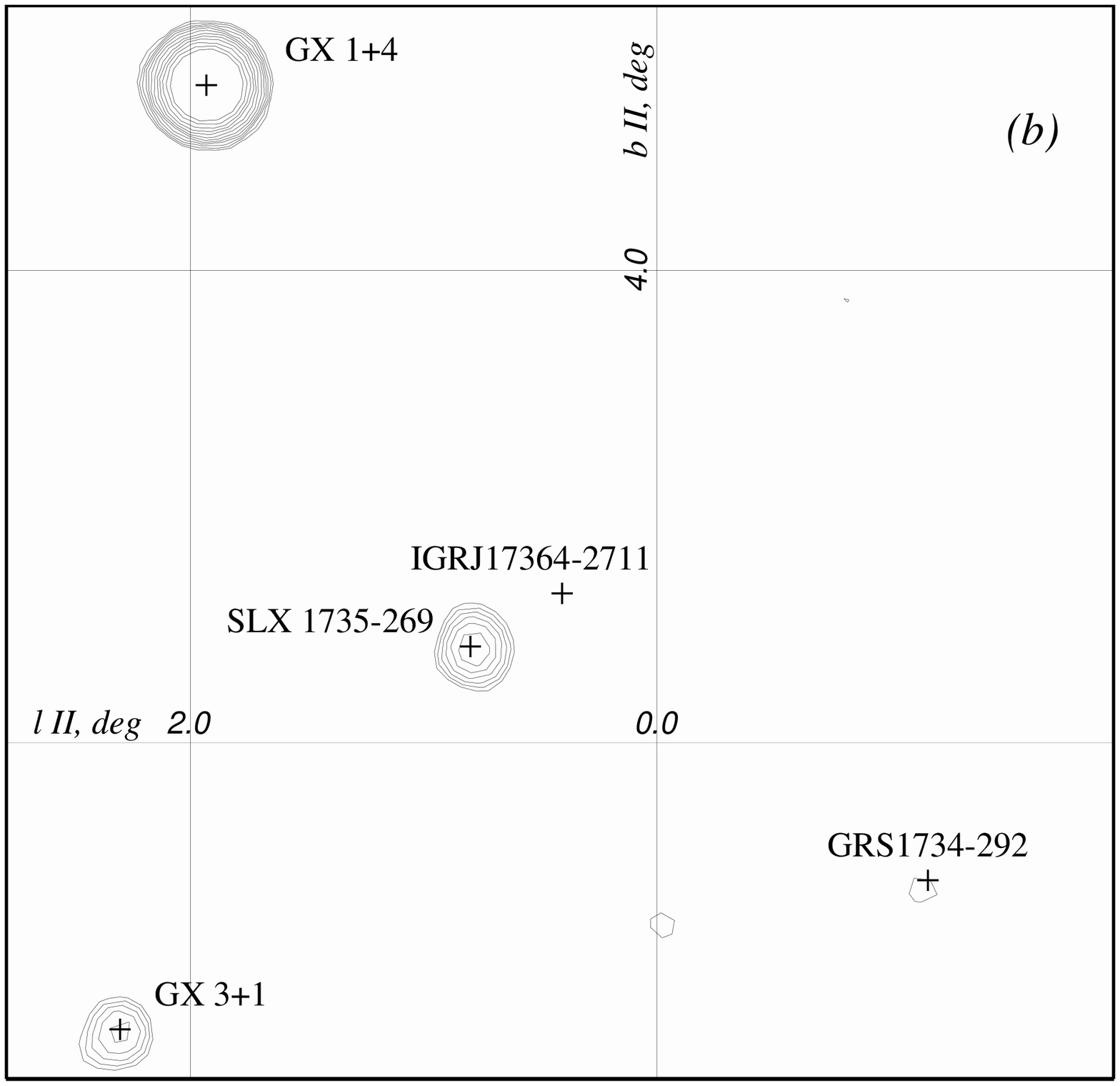,width=0.33\linewidth}
\epsfig{file=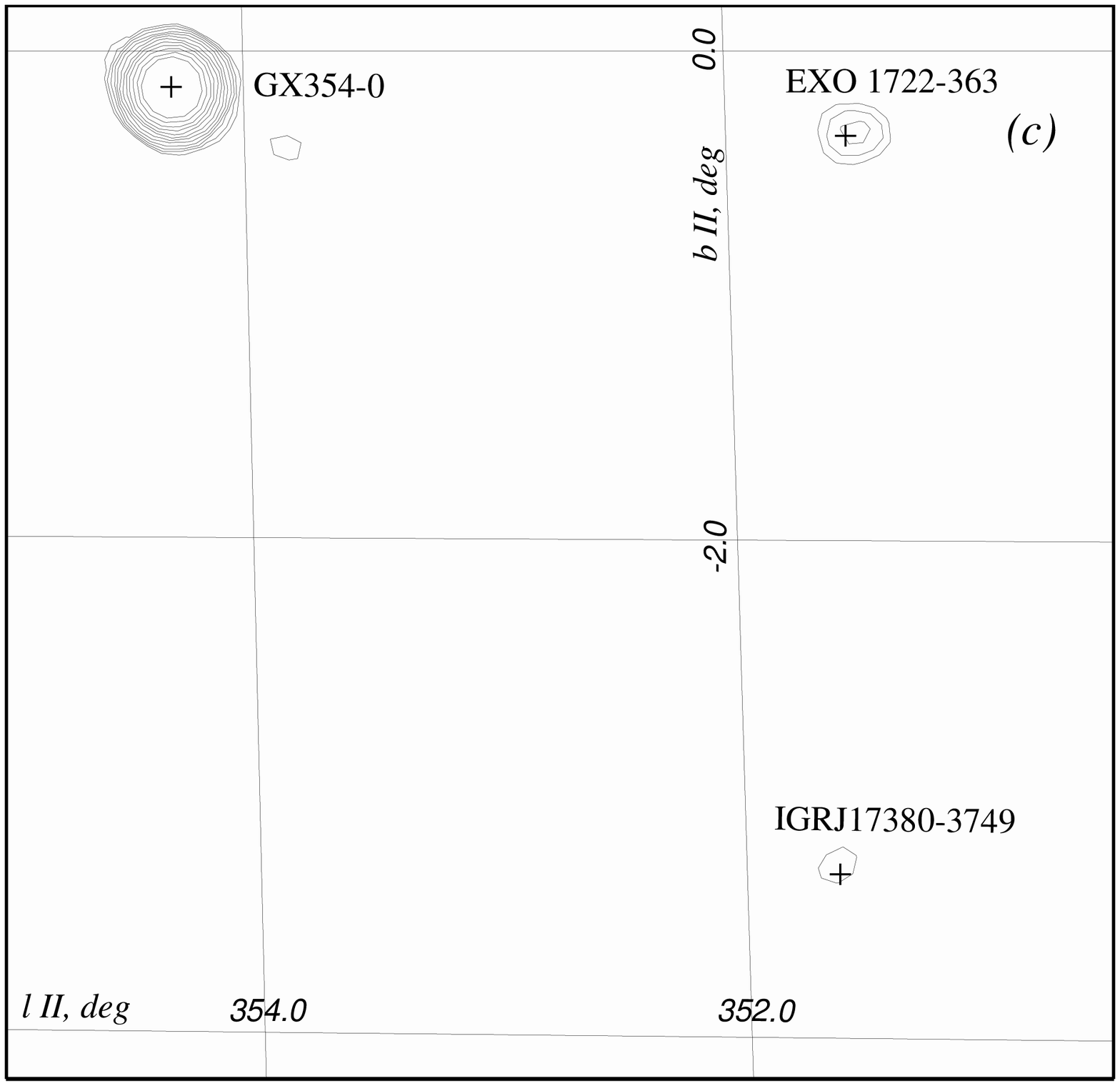,width=0.33\linewidth}
\caption{X-ray images obtained with IBIS/ISGRI on February
16--17, 2004: (a) during 15 s of the burst detected from the new
burster \mbox{IGR J17364-2711} (15-25 keV), (b) and (c) during
the entire observing session except the burst interval (165 ks,
18--45 keV).  Contours show regions of confident detection of
sources at the $S/N$ levels of\/ $3.5, 4.4, 5.5, 6.8, 8.5, ...,
40$.\label{fig2}}
\end{figure*}

responding to an individual pointing) with a time resolution of
5 s. More than 32800 individual observations were analyzed for the
presence of bursts with a total exposure time of over 67 Ms. An
excess of the signal-to-noise ratio $(S-\overline{S})/N$ above
the preset threshold $s_0=5.25$ in a particular time bin served
a criterion for a burst. Since the number of counts in each bin
obeys a Poisson distribution, there is a low, but finite
probability $p_0\simeq8\times10^{-8}\!$ (for
$s_0\simeq5.25$)$\,$of$\,$recording a$\,$random spike even in
the$\,$absence of a real burst.
The selected threshold $s_0$ ensures that only one such spike 
in the entire time series (with $M\sim 1.3\times10^7$ bins) may be
recorded ($p_0\times M\simeq 1$). Since the
total count rate of the detector depended on the emission from
all sources within the IBIS field of view (FOV), the mean count
rate $\overline{S}$ and the noise level
$N=(\overline{S^2}-\overline{S}^2)^{1/2}$ were determined
independently for each pointing. For all the detected
bursts, we reconstructed the images of the sky area within the
IBIS FOV (the IMA phase of OSA 4.2) accumulated
with the same exposure at the burst time and immediately before
the burst and compared the statistical significance of 
sources detected in them to reveal the burst source.

\section{\bf RESULTS}
There were 1788 bursts found in the IBIS/ISGRI count rate
histograms; the sources of 319 bursts were also detected in the
IBIS images. Some of them were associated with cosmic GRBs (11
events), activity of SGR 1806-20 (59) or several bright XBs
(32). The other 217 of the localized bursts (see Table) were
associated with 15 known and 2 new X-ray bursters (138 of them
--- with \mbox{GX 354-0}). Among the new bursters one was
actually a weak unidentified source AX\,J1754.2-2754 discovered
by ASCA in 1999 (note that 1A\,1246-588, one of the 15 known
sources, was identified as a burster only a few weeks before
this Workshop --- due to an intense type-I burst detected from
its positon by RXTE \citep{kuulkers06}). An X-ray burst from
the second new burster was detected on February 17, 2004, when
IBIS observed the Galactic center region. Because of peculiarity
of the IBIS coded mask, the burster position was determined with
duality \citep{chelovekov06}: it could be either R.A.=$17\uh36\um28\us$,
Decl.=$-27\deg11\arcmin56\arcsec$ (see Fig.\,\ref{fig2}a) or
R.A.=$17\uh38\um05\us$, Decl.=$-37\deg49\arcmin05\arcsec$ (epoch
2000.0, error radius for the positions is 2\arcmin). The $S/N$
ratio was the same ($\simeq8.9$) for both the positions, however
in the first case this ratio was reached with the larger number
of counts (394) compared to the second case (316).\,Taking this
fact into account we named the source \mbox{IGR\,J17364-2711},
although the second name, IGR\,J17380-3749, cannot be completely
rejected. As Fig.\,\ref{fig2}b shows IGR\,J17364-2711 was not
seen in the image accumulated during the entire observing
session of \mbox{February\,16--17,} except the pointing during
which the burst occurred. Fig.\,\ref{fig2}c shows another part
of the same image with an excess near the position of
IGR\,J17380-3749. It is however too weak ($S/N\simeq4.3$) to
ensure the presence of a real persistent source.

The photon flux measured during $\Delta T\simeq 1$ s at the
burst maximum reached $1.6\pm0.3$ Crab, that corresponded to a
15--25 keV luminosity $\simeq 8\times10^{37}$ \ergs\ at the
Galactic center distance of $8.5$ kpc. The burst spectrum was
very soft and the flux at energies $\ga30$ keV fell below the
detection level. Assuming the spectrum to have a Wien shape with
$kT\simeq2.5-2.8$ keV typical of bursts with photospheric
expansion, we found the bolometric luminosity of this burst,
$L_{\rm B}\simeq (4-5)\times10^{38}$ \ergs\ which is actually
very close to the Eddington one. It was not possible to refine
$kT$ from observations in the standard X-ray band (with the
INTEGRAL/JEM-X monitor) since the burst occurred at the very edge
of the JEM-X FOV, narrower than the IBIS one, and was not detected.

The $3\sigma$ limit on the persistent 18--45 keV flux from
IGR\,J17364-2711 was 3 mCrab. Assuming a power-law spectrum with
the photon index $\gamma\simeq2.1$ for the source, we obtain a
fairly stringent limit on its 2--45 keV luminosity, $L_{\rm
X}\la1.5\times10^{36}$ \ergs. Thus, IGR\,J17364-2711 may complement
the list of X-ray bursters that have never been observed in a state of
persistent X-ray emission.

\begin{table*}
\begin{minipage}[t]{\columnwidth}
\section*{ACKNOWLEDGMENTS}
This research was supported by the Russian Foundation for Basic
Research (project 05-02-17454), the Presidium of the Russian
Academy of Sciences (the ``Origin and evolution of stars and
galaxies'' program), and the Program of the Russian President
for Support of Leading Scientific Schools (project
NSh-1100.2006.2).
\end{minipage}\hspace{\columnsep}\begin{minipage}[t]{\columnwidth}

\end{minipage}
\centering

\vspace{0.3cm}

{{\it Table.} \rm X-ray bursts detected by the IBIS/ISGRI
telescope in the 15-25 keV band and their peak fluxes (in Crab units)\label{tab:burst_cat}}

\vspace{0.2cm}

\begin{tabular}{rlcccp{4pt}rlccc}
\hline
\hline
\\ [-3.5mm]
{\#}&
{Source}&
\multicolumn{3}{c}{Burst maximum ($\Delta T\simeq1$ s)}&&
{\#}&
{Source}&
\multicolumn{3}{c}{Burst maximum ($\Delta T\simeq1$ s)}\\
{}&
{}&
{UTC}&
{h:m:s}&
{Crab}&&
{}&
{}&
{UTC}&
{h:m:s}&
{Crab}\\ \hline
\\ [-3.5mm]
$ 1$&       GX~354-0&28.02.03&07:55:06&$2.52$ &&$50$&       GX~354-0&13.09.03&16:40:43&$1.03$ \\
$ 2$&       GX~354-0&01.03.03&00:04:50&$2.48$ &&$51$&       GX~354-0&13.09.03&22:28:39&$2.97$ \\
$ 3$&       GX~354-0&01.03.03&16:05:33&$3.04$ &&$52$&       GX~354-0&14.09.03&15:02:23&$2.72$ \\
$ 4$&       GX~354-0&02.03.03&07:42:22&$3.04$ &&$53$&       GX~354-0&14.09.03&20:55:10&$2.17$ \\
$ 5$&    4U~1636-536&04.03.03&19:18:02&$1.13$ &&$54$&       GX~354-0&15.09.03&09:40:18&$1.71$ \\
$ 6$&    4U~1702-429&09.03.03&21:51:13&$2.93$ &&$55$&       GX~354-0&15.09.03&15:49:09&$3.10$ \\
$ 7$&    4U~1608-522&09.03.03&22:35:05&$2.57$ &&$56$&   SLX~1735-269&15.09.03&17:42:29&$2.16$ \\
$ 8$&       GX~354-0&12.03.03&10:22:26&$2.70$ &&$57$&       GX~354-0&17.09.03&02:42:50&$2.79$ \\
$ 9$&    4U~1702-429&12.03.03&11:11:03&$3.09$ &&$58$&       GX~354-0&17.09.03&08:58:31&$2.95$ \\
$10$&    4U~1608-522&13.03.03&13:49:37&$2.61$ &&$59$&       GX~354-0&18.09.03&10:33:36&$2.42$ \\
$11$&    4U~1702-429&15.03.03&02:38:07&$2.44$ &&$60$&       GX~354-0&19.09.03&16:12:10&$3.18$ \\
$12$&    4U~1702-429&15.03.03&18:22:49&$1.74$ &&$61$&       GX~354-0&20.09.03&05:40:37&$3.70$ \\
$13$&       GX~354-0&15.03.03&20:36:46&$1.22$ &&$62$&       GX~354-0&20.09.03&23:47:03&$3.44$ \\
$14$&       GX~354-0&03.04.03&08:40:18&$3.10$ &&$63$&       GX~354-0&21.09.03&14:07:40&$2.93$ \\
$15$&        Aql~X-1&06.04.03&07:42:15&$1.42$ &&$64$&       GX~354-0&22.09.03&17:38:26&$2.81$ \\
$16$&    4U~1724-307&06.04.03&18:32:31&$1.26$ &&$65$&       GX~354-0&23.09.03&02:16:11&$2.40$ \\
$17$&       GX~354-0&06.04.03&19:45:29&$1.54$ &&$66$&   SLX~1735-269&23.09.03&05:11:43&$1.04$ \\
$18$&       GX~354-0&07.04.03&03:26:31&$1.78$ &&$67$&       GX~354-0&23.09.03&10:53:38&$2.37$ \\
$19$&    4U~1636-536&11.04.03&18:13:18&$0.52$ &&$68$&       GX~354-0&23.09.03&18:15:09&$2.61$ \\
$20$&    4U~1702-429&15.04.03&06:47:16&$2.54$ &&$69$&   SLX~1735-269&23.09.03&23:13:11&$1.05$ \\
$21$&     4U~1812-12&21.04.03&03:36:36&$2.55$ &&$70$&       GX~354-0&24.09.03&03:52:12&$3.18$ \\
$22$&     4U~1812-12&25.04.03&10:54:25&$3.60$ &&$71$&       GX~354-0&24.09.03&11:01:26&$3.00$ \\
$23$&    2S~0918-549&16.06.03&20:09:13&$3.61$ &&$72$&SAX~J1712.6-3739&24.09.03&14:00:09&$2.18$\\
$24$&    4U~1702-429&18.08.03&10:05:10&$2.46$ &&$73$&       GX~354-0&24.09.03&18:20:21&$2.63$ \\
$25$&       GX~354-0&23.08.03&16:14:02&$0.89$ &&$74$&    4U~1608-522&26.09.03&02:38:55&$3.73$ \\
$26$&       GX~354-0&24.08.03&22:20:44&$1.69$ &&$75$&    4U~1608-522&26.09.03&15:34:51&$3.79$ \\
$27$&SAX~J1712.6-3739&25.08.03&18:45:43&$1.78$&&$76$&    4U~1608-522&27.09.03&05:10:24&$4.95$ \\
$28$&       GX~354-0&27.08.03&19:59:14&$1.82$ &&$77$&     4U~1812-12&27.09.03&16:08:45&$3.06$ \\
$29$&       GX~354-0&28.08.03&01:24:04&$0.91$ &&$78$&       GX~354-0&04.10.03&22:06:42&$1.68$ \\
$30$&       GX~354-0&28.08.03&06:01:30&$1.86$ &&$79$&       GX~354-0&05.10.03&09:34:42&$2.72$ \\
$31$&       GX~354-0&29.08.03&14:31:29&$2.01$ &&$80$&       GX~354-0&08.10.03&09:58:37&$1.03$ \\
$32$&       GX~354-0&29.08.03&19:23:36&$2.19$ &&$81$&       GX~354-0&17.02.04&04:47:50&$3.27$ \\
$33$&       GX~354-0&31.08.03&15:54:18&$1.13$ &&$82$&IGR~J17364-2711&17.02.04&14:41:30&$1.59$ \\
$34$&       GX~354-0&03.09.03&03:26:34&$1.62$ &&$83$&       GX~354-0&19.02.04&21:06:44&$2.13$ \\
$35$&       GX~354-0&03.09.03&08:39:32&$1.49$ &&$84$&       GX~354-0&20.02.04&02:44:00&$2.11$ \\
$36$&       GX~354-0&03.09.03&18:02:39&$1.08$ &&$85$&       GX~354-0&20.02.04&12:00:23&$1.13$ \\
$37$&     4U~1812-12&06.09.03&00:23:32&$4.04$ &&$86$&       GX~354-0&27.02.04&10:55:16&$3.08$ \\
$38$&       GX~354-0&07.09.03&20:30:07&$1.55$ &&$87$&       GX~354-0&27.02.04&13:32:36&$1.53$ \\
$39$&       GX~354-0&08.09.03&13:41:36&$2.66$ &&$88$&       GX~354-0&27.02.04&15:32:03&$2.11$ \\
$40$&    4U~1724-307&08.09.03&18:48:30&$1.48$ &&$89$&       GX~354-0&02.03.04&07:34:38&$1.52$ \\
$41$&       GX~354-0&08.09.03&19:41:21&$2.63$ &&$90$&         GX~3+1&02.03.04&09:25:34&$0.98$ \\
$42$&       GX~354-0&09.09.03&03:11:36&$2.95$ &&$91$&    4U~1724-307&03.03.04&04:14:60&$1.40$ \\
$43$&       GX~354-0&09.09.03&16:28:54&$1.45$ &&$92$&       GX~354-0&08.03.04&04:14:45&$1.03$ \\
$44$&       GX~354-0&09.09.03&22:22:24&$2.58$ &&$93$&    4U~1608-522&20.03.04&20:59:32&$3.72$ \\
$45$&       GX~354-0&11.09.03&05:04:28&$3.92$ &&$94$&    4U~1608-522&21.03.04&01:03:47&$1.83$ \\
$46$&       GX~354-0&11.09.03&10:57:50&$1.99$ &&$95$&        Aql~X-1&24.03.04&17:03:35&$1.58$ \\
$47$&       GX~354-0&11.09.03&21:59:51&$2.92$ &&$96$&       GX~354-0&29.03.04&02:40:47&$0.99$ \\
$48$&       GX~354-0&12.09.03&03:12:25&$2.04$ &&$97$&       GX~354-0&30.03.04&03:25:33&$1.25$ \\
$49$&       GX~354-0&12.09.03&09:22:40&$2.58$ &&$98$&   SLX~1744-299&30.03.04&03:37:46&$0.81$ \\
\hline					    
\end{tabular}			    
\end{table*}

\begin{table*}
\vspace{-0.5mm}
\centering
%

\begin{tabular}{rlcccp{4pt}rlccc}
\hline
\hline
\\ [-3.6mm]
{\#}&
{Source}&
\multicolumn{3}{c}{Burst maximum ($\Delta T\simeq1$ s)}&&
{\#}&
{Source}&
\multicolumn{3}{c}{Burst maximum ($\Delta T\simeq1$ s)}\\
{}&
{}&
{UTC}&
{h:m:s}&
{Crab}&&
{}&
{}&
{UTC}&
{h:m:s}&
{Crab}\\ \hline
\\ [-3.6mm]
 $99$&   KS~1741-293&30.03.04&03:43:45&$0.88$ &&$159$&    4U~1702-429&25.02.05&13:58:35&$1.80$\\
$100$&      GX~354-0&31.03.04&03:09:05&$0.85$ &&$160$&    4U~1702-429&27.02.05&16:08:41&$2.47$\\
$101$&      GX~354-0&01.04.04&23:36:53&$0.99$ &&$161$&    4U~1608-522&03.03.05&19:46:16&$3.80$\\
$102$&       Aql~X-1&28.04.04&07:54:48&$1.96$ &&$162$&    4U~1608-522&04.03.05&13:13:59&$2.09$\\
$103$&       Aql~X-1&01.05.04&22:56:47&$1.89$ &&$163$&    4U~1608-522&05.03.05&22:37:27&$2.56$\\
$104$&      GX~354-0&22.08.04&14:37:53&$1.75$ &&$164$&    4U~1636-536&06.03.05&01:25:52&$1.37$\\
$105$&  SLX~1735-269&23.08.04&17:23:59&$1.04$ &&$165$&    4U~1608-522&06.03.05&12:30:56&$4.21$\\
$106$&      GX~354-0&23.08.04&21:53:59&$1.59$ &&$166$&    4U~1608-522&06.03.05&22:55:52&$5.02$\\
$107$&      GX~354-0&01.09.04&01:22:21&$0.77$ &&$167$&    4U~1608-522&08.03.05&07:56:40&$5.07$\\
$108$&      GX~354-0&01.09.04&15:25:02&$1.65$ &&$168$&    4U~1608-522&09.03.05&14:08:27&$4.27$\\
$109$&      GX~354-0&01.09.04&19:26:43&$1.61$ &&$169$&       GX~354-0&17.03.05&08:16:32&$1.11$\\
$110$&      GX~354-0&01.09.04&23:12:18&$1.44$ &&$170$&       GX~354-0&18.03.05&00:53:47&$2.52$\\
$111$&      GX~354-0&02.09.04&03:23:25&$2.80$ &&$171$&       GX~354-0&18.03.05&23:12:29&$2.80$\\
$112$&      GX~354-0&02.09.04&07:16:21&$1.65$ &&$172$&       GX~354-0&20.03.05&19:38:07&$2.50$\\
$113$&      GX~354-0&03.09.04&14:32:30&$2.97$ &&$173$&       GX~354-0&20.03.05&23:28:19&$2.33$\\
$114$&      GX~354-0&03.09.04&18:39:35&$1.83$ &&$174$&       GX~354-0&21.03.05&03:40:50&$2.45$\\
$115$&      GX~354-0&03.09.04&23:17:42&$1.66$ &&$175$&SAX~J1712.6-3739&21.03.05&19:06:05&$2.57$\\ 
$116$&      GX~354-0&04.09.04&09:06:02&$3.40$ &&$176$&       GX~354-0&22.03.05&01:24:37&$2.33$\\
$117$&      GX~354-0&04.09.04&16:19:11&$0.84$ &&$177$&       GX~354-0&22.03.05&23:59:45&$2.62$\\
$118$&      GX~354-0&04.09.04&23:50:03&$3.55$ &&$178$&       GX~354-0&23.03.05&11:44:33&$1.68$\\
$119$&   4U~1702-429&07.09.04&10:51:47&$2.26$ &&$179$&       GX~354-0&24.03.05&15:39:10&$1.75$\\ 
$120$&      GX~354-0&07.09.04&11:09:51&$1.69$ &&$180$&       GX~354-0&24.03.05&21:14:22&$2.43$\\ 
$121$&      GX~354-0&07.09.04&14:27:14&$3.01$ &&$181$&       GX~354-0&25.03.05&02:22:48&$2.53$\\
$122$&      GX~354-0&07.09.04&18:04:56&$1.03$ &&$182$&       GX~354-0&25.03.05&23:51:29&$2.45$\\ 
$123$&      GX~354-0&07.09.04&21:30:48&$1.29$ &&$183$&       GX~354-0&26.03.05&09:05:20&$3.08$\\
$124$&      GX~354-0&08.09.04&01:54:26&$2.61$ &&$184$&       GX~354-0&27.03.05&08:45:39&$2.00$\\
$125$&      GX~354-0&08.09.04&09:00:22&$1.94$ &&$185$&    4U~1702-429&27.03.05&11:49:04&$1.98$\\ 
$126$&      GX~354-0&08.09.04&12:41:39&$1.79$ &&$186$&       GX~354-0&27.03.05&15:23:04&$2.21$\\
$127$&      GX~354-0&08.09.04&15:50:10&$3.27$ &&$187$&       GX~354-0&28.03.05&01:47:04&$0.73$\\ 
$128$&   4U~1702-429&08.09.04&17:43:26&$1.05$ &&$188$&     H~0614+091&31.03.05&07:13:08&$8.43$\\ 
$129$&   4U~1636-536&11.09.04&04:16:54&$1.36$ &&$189$&       GX~354-0&04.04.05&01:17:29&$1.39$\\ 
$130$&      GX~354-0&15.09.04&12:56:11&$1.45$ &&$190$&    4U~1702-429&04.04.05&04:55:51&$2.92$\\
$131$&      GX~354-0&19.09.04&12:00:32&$1.39$ &&$191$&    4U~1702-429&06.04.05&08:56:54&$1.81$\\ 
$132$&      GX~354-0&22.09.04&17:37:15&$1.44$ &&$192$&       GX~354-0&07.04.05&10:06:45&$0.90$\\
$133$&      GX~354-0&23.09.04&04:10:50&$1.56$ &&$193$&    4U~1702-429&08.04.05&06:04:19&$2.04$\\
$134$&      GX~354-0&23.09.04&07:37:29&$0.97$ &&$194$&    4U~1702-429&10.04.05&03:01:03&$2.45$\\
$135$&      GX~354-0&30.09.04&11:47:22&$1.74$ &&$195$&       GX~354-0&10.04.05&07:42:23&$1.04$\\
$136$&      GX~354-0&30.09.04&14:54:25&$2.50$ &&$196$&       GX~354-0&14.04.05&06:33:29&$1.33$\\
$137$&      GX~354-0&01.10.04&03:11:27&$0.45$ &&$197$&AX~J1754.2-2754&16.04.05&22:11:04&$1.36$\\
$138$&      GX~354-0&01.10.04&06:53:21&$2.27$ &&$198$&    4U~1724-307&16.04.05&22:17:44&$1.15$\\
$139$&      GX~354-0&01.10.04&14:28:34&$2.60$ &&$199$&     4U~1812-12&28.04.05&06:00:38&$4.65$\\ 
$140$&      GX~354-0&01.10.04&22:11:34&$2.60$ &&$200$&    1A~1246-588&27.06.05&11:06:54&$2.99$\\
$141$&      GX~354-0&02.10.04&01:58:58&$2.45$ &&$201$&       GX~354-0&12.08.05&16:47:26&$1.11$\\
$142$&      GX~354-0&02.10.04&05:59:15&$2.46$ &&$202$&    4U~1608-522&12.08.05&22:09:58&$3.99$\\
$143$&      GX~354-0&02.10.04&10:12:10&$1.71$ &&$203$&       GX~354-0&17.08.05&03:06:31&$2.16$\\
$144$&      GX~354-0&02.10.04&14:16:06&$3.06$ &&$204$&    4U~1608-522&17.08.05&08:55:27&$3.67$\\
$145$&      GX~354-0&02.10.04&21:50:09&$3.05$ &&$205$&       GX~354-0&25.08.05&16:19:41&$2.50$\\
$146$&    4U~1812-12&04.10.04&03:15:49&$4.29$ &&$206$&    4U~1636-536&25.08.05&21:28:22&$0.89$\\
$147$&      GX~354-0&16.10.04&07:27:17&$1.20$ &&$207$&    4U~1702-429&26.08.05&07:07:35&$2.59$\\
$148$&      GX~354-0&17.10.04&10:58:54&$1.25$ &&$208$&    4U~1608-522&26.08.05&10:51:25&$5.11$\\
$149$&      GX~354-0&20.10.04&06:37:13&$1.05$ &&$209$&    4U~1608-522&27.08.05&11:34:03&$4.60$\\
$150$&      GX~354-0&20.10.04&11:35:45&$1.35$ &&$210$&       GX~354-0&28.08.05&13:24:16&$2.90$\\
$151$&   4U~1702-429&18.02.05&15:35:56&$1.92$ &&$211$&       GX~354-0&28.08.05&21:39:42&$2.27$\\
$152$&   4U~1702-429&19.02.05&02:26:13&$2.03$ &&$212$&       GX~354-0&29.08.05&08:29:35&$2.10$\\
$153$&   4U~1702-429&19.02.05&14:27:24&$2.84$ &&$213$&    4U~1608-522&29.08.05&10:39:19&$4.52$\\
$154$&   4U~1636-536&19.02.05&15:38:52&$1.34$ &&$214$&    4U~1636-536&29.08.05&13:47:36&$0.82$\\
$155$&      GX~354-0&21.02.05&10:06:19&$1.17$ &&$215$&    4U~1702-429&30.08.05&06:21:50&$3.00$\\
$156$&   4U~1636-536&22.02.05&13:48:56&$1.14$ &&$216$&       GX~354-0&30.08.05&12:10:38&$4.62$\\
$157$&   4U~1702-429&24.02.05&20:22:07&$1.13$ &&$217$&    4U~1702-429&30.08.05&16:45:34&$1.32$\\
$158$&   4U~1702-429&25.02.05&04:53:55&$2.27$ &&\\
\hline
\end{tabular}					
\end{table*}					
\end{document}